\documentclass[aps,prl,twocolumn,amsmath,amssymb]{revtex4}
\usepackage{graphicx}
\usepackage{dcolumn}
\usepackage{bm}

\begin{document}

\title{Impact Excitation by Hot Carriers in Carbon Nanotubes}

\author{Vasili Perebeinos$*$ and Phaedon Avouris}
\affiliation{IBM Research Division, T. J. Watson Research Center,
Yorktown Heights, New York 10598}

\date{\today}

\begin{abstract}
We investigate theoretically the efficiency of intra-molecular hot
carrier induced impact ionization and excitation processes in carbon
nanotubes. The electron confinement and reduced screening lead to
drastically enhanced excitation efficiencies over those in bulk
materials. Strong excitonic coupling favors neutral excitations over
ionization, while the impact mechanism populates a different set of
states than that produced by photoexcitation. The excitation rate is
strongly affected by optical phonon excitation and a simple scaling
of the rate with the field strength and optical phonon temperature
is obtained.
\end{abstract}

\maketitle

The excellent electrical properties and direct gap of carbon
nanotubes (CNTs) offer the possibility of a unified electronic and
optoelectronic technology based on this material \cite{Avouris}. The
electroluminescence (EL) in CNTs field effect transistors has
already been demonstrated, where electrons and holes were
independently injected in an ambipolar device
\cite{Misewich,Freitag1,Freitag2}. Recently, light emission has been
observed under unipolar transport conditions in the suspended tubes
\cite{Chen} and from the CNT-metal contacts \cite{Marty,Freitag2}.
In the suspended sample the light intensity is stronger by a factor
of $\sim$1000 compared to ambipolar devices. The mechanism of this
enhanced EL was proposed to involve an unusually efficient
intra-nanotube impact excitation process involving the hot carriers
in CNTs, generated by the high local fields at the interface between
the suspended and nonsuspended parts of the tube and the
metal-nanotube contact. By utilizing such impact excitation process
in quasi one-dimensional confined structures, high exciton densities
can be achieved for probing the interactions of such boson systems
and for generating ultra bright nanoscale light sources used in
nanophotonics.

Here, we provide detailed theoretical study of the impact excitation
in semiconducting CNTs. In low dimensional materials, the Coulomb
interactions are weakly screened, and large exciton binding energies
in CNTs have been predicted theoretically
\cite{Ando,Spataru,Perebeinos1} and verified experimentally
\cite{Wang,Germans,Qiu}. This presents an additional challenge for
the computation of impact excitation rates in CNTs. We find that the
impact scattering rate is very fast, up to five orders of magnitude
larger than those in bulk semiconductors \cite{Kuligk}. This can be
easily rationalized since the same Coulomb interactions responsible
for the large exciton binding are also responsible for the impact
excitation scattering. The electron-hole correlations in the final
state make the impact excitation process, on the average, twice more
efficient than the impact ionization, where the created electron
hole pairs are treated as independent particles. Impact excitation
generates a different initial distribution of excited states than
that produced by photon excitation due to the different selection
rules and momentum conservation requirements.

It has recently been shown that under high bias the energetic
carriers in nanotubes excite optical and zone-boundary phonons
\cite{Yao,Javey,Park,Perebeinos3} and generate a non-equilibrium
phonon distribution \cite{Lazzeri,Dai}, particularly when energy
dissipation to the substrate is suppressed as in suspended CNTs
\cite{Dai}. We have calculated the effect of optical phonon
excitation and found that the exciton production rate depends
exponentially both on the hot phonon temperature and on the electric
field, and the effect can be described by a simple scaling relation.
As a result, optical phonon excitation significantly increases the
EL intensity under unipolar transport conditions. The energy spectra
of the excitons generated consist of several peaks corresponding to
excitons with different angular momenta $L$: $E^{2}_{11}$,
$E^{3}_{12}$, and $E^4_{22}$, where $E^{L}_{ij}$ stands for exciton
with electron and hole primarily from bands $i$ and $j$. The
relative intensities and the overall excitation efficiencies are
functions of the electric field, the optical phonon temperature, and
the CNT chirality.

The impact ionization scattering rate of an electron with momentum
and energy ($k_1$, $\varepsilon^c_{k_1}$) scattered to state
($k_1-q$, $\varepsilon^c_{k_1-q}$)  plus an e-h pair with a hole
($-k_2$, -$\varepsilon^v_{k_2}$) and an electron ($k_2+q$,
$\varepsilon^c_{k_2+q}$)  is given by \cite{Landsberg}:
\begin{eqnarray}
W_{k_1,k_1-q}&=&\frac{2\pi}{\hbar}\sum_{k_2}\left\vert
M_{\nu}\right\vert^2\delta\left(\varepsilon^c_{k_1}-\varepsilon^v_{k_2}-
\varepsilon^c_{k_1-q}-\varepsilon^c_{k_2+q}\right) \nonumber \\
\vert M_{\nu}\vert^2&=&\vert M^{d}_{\nu}-M^e_{\nu}\vert^2+\vert
M^{d}_{\nu}\vert^2+\vert M^{e}_{\nu}\vert^2 \label{eq1}
\end{eqnarray}
where ${\nu}=k_1k_2q$,  $M^d_{\nu}=J(k_1k_2,k_2+qk_1-q)$ and
$M^e_{\nu}=J(k_1k_2,k_1-qk_2+q)$ are the direct and the exchange
integrals:
\begin{eqnarray}
&&J(k_1k_2,k_1-qk_2+q)=\int d\vec{r}_1 d\vec{r}_2
\Psi^{*}_{ck_1}(\vec{r}_1)\Psi^*_{vk_2}(\vec{r}_2) \nonumber
\\
&&\Psi^{}_{ck_1-q}(\vec{r}_1)\Psi^{}_{ck_2+q}(\vec{r}_2)
V(\vec{r}_1,\vec{r}_2) \label{eq2}
\end{eqnarray}
$V(\vec{r}_1,\vec{r}_2)$ is the Coulomb potential screened by the
dielectric environment of the nanotube \cite{Perebeinos1} and
$\Psi^{}_{c}(\vec{r})$ and $\Psi^{*}_{v}(\vec{r})$ are the
electron and the hole wavefunctions, given by a $\pi$-orbital
tight-binding model with hopping matrix element $t=3.0$ eV. In
Eq.~(\ref{eq1}) we neglect e-h pair occupation in the initial
state and, hence, processes where electron can absorb an e-h pair.

To include the effect of e-h correlation in the final state we
follow \cite{Chang} and show that the impact excitation scattering
rates due to the production of singlet (S) and triplet (T)
excitons are given by:
\begin{eqnarray}
W^{S}_{k_1k_1-q}&=&\frac{\pi}{\hbar}\sum_{\mu}\left\vert
\tilde{M}^S_{k_1q\mu}\right\vert^2
\delta\left(\varepsilon^c_{k_1}-
\varepsilon^c_{k_1-q}-E^S_{q\mu}\right) \nonumber \\
W^{T}_{k_1k_1-q}&=&\frac{\pi}{\hbar}\sum_{\mu}\left\vert
\tilde{M}^T_{k_1q\mu}\right\vert^2\delta\left(\varepsilon^c_{k_1}-
\varepsilon^c_{k_1-q}-E^T_{q\mu}\right) \nonumber \\
\left\vert \tilde{M}^{S}_{k_1q\mu}\right\vert^2&=&\left\vert
\tilde{M}^d_{k_1q\mu}-\tilde{M}^e_{k_1q\mu}\right\vert^2+\left\vert\tilde{
M}^d_{k_1q\mu}\right\vert^2 \nonumber \\
\left\vert \tilde{M}^{T}_{k_1q\mu}\right\vert^2&=&\left\vert
\tilde{M}^d_{k_1q\mu}-\tilde{M}^e_{k_1q\mu}\right\vert^2+\left\vert
\tilde{M}^d_{k_1q\mu}\right\vert^2+2\left\vert
\tilde{M}^e_{k_1q\mu}\right\vert^2
\nonumber \\
\tilde{M}^d_{k_1q\mu}&=&\sum_{k2}A_{k_2q}^{\mu}M^d_{k_1k_2q}
\nonumber \\
\tilde{M}^e_{k_1q\mu}&=&\sum_{k2}A_{k_2q}^{\mu}M^e_{k_1k_2q}
 \label{eq4}
\end{eqnarray}
where $E_{q\mu}^S$ and $E_{q\mu}^T$ are the singlet and triplet
exciton transition energies \cite{note1}, $A_{kq}^{\mu}$ is the
solution of the Bethe-Salpeter equation \cite{Spataru,Perebeinos1}
for the exciton two-particle wavefunction:
\begin{eqnarray}
\Phi_q^{\mu}(\vec{r}_1,\vec{r}_2)=\sum_k
A_{kq}^{\mu}\Psi^{}_{ck+q}(\vec{r}_1)\Psi^{*}_{vk}(\vec{r}_2).
 \label{eq3}
\end{eqnarray}
We model the screening of the Coulomb interaction by the dielectric
constant of the medium embedding the nanotube \cite{Perebeinos1},
both for the Bethe-Salpeter equation kernel and the impact
ionization Coulomb potential Eq.~(\ref{eq2}). The CNT diameter
dependence of exciton binding energies obtained by two-photon
florescence excitation spectroscopy \cite{Dukovic} agree very well
with our model calculations \cite{Perebeinos1} by choosing
$\epsilon=3.3$, a value which we use in the rest of the paper.

\begin{figure}
\includegraphics[height=2.4in]{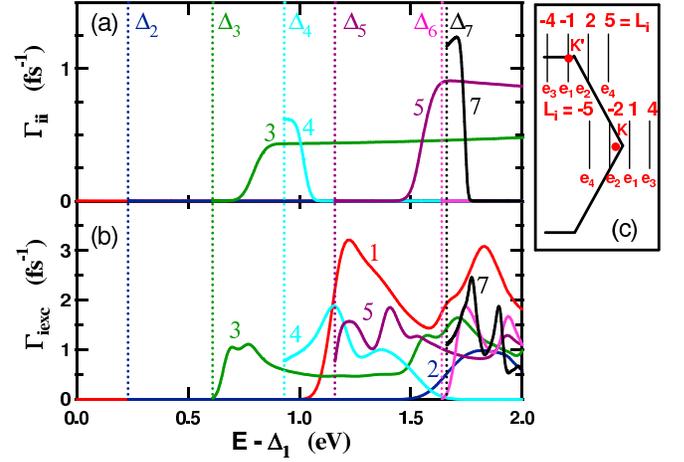}
\caption{\label{fig1}  (a) Impact ionization rates and (b) impact
excitation rates for a (25,0) nanotube ($d=2.0$ nm) as a function of
energy (measured from the bottom of the first conduction band
$\Delta_1$), for the first seven conduction bands shown in red,
blue, green, cyan, magenta, light red, and black in increasing order
of the band index. The vertical dashed lines correspond to the
bottoms of the conduction bands $\Delta_i$ ($i=2, 3... 7$) relative
to $\Delta_1$. (c) The vertical lines are allowed k-points of
zig-zag tube with $mod(n,3)=1$ for the first four doubly degenerate
bands $e_i^{L_i}$. The angular momenta $L_i$, in units of $2/3d$,
measures a minimum distance from the $K$ ($K'$) point of graphene to
the 1D lines of allowed k-points of a nanotube.}
\end{figure}

The impact ionization and impact excitation scattering rates for a
tube of diameter $d=2$ nm are shown in Fig.~\ref{fig1}a and
\ref{fig1}b respectively. This is a typical diameter of tubes used
in opto-electronic applications and grown by chemical vapor
deposition. The impact excitation rates in CNTs are roughly twice as
large as the impact ionization rates due to the e-h confinement in
the final state. To understand the structure of the impact
ionization scattering rates in Fig.~\ref{fig1}a, we need to consider
the implications of energy and momentum conservation. The momentum
has two components: a discrete angular momentum, which labels the
bands, and a linear momentum along the tube axis. The CNT bands are
non-parabolic and an angular momentum conservation law plays a
crucial role in determining the threshold energy of the impact
excitation \cite{footnote1}. As seen from Fig.~\ref{fig1}c, angular
momentum conservation allows electrons in the third, $e_3$, and the
forth, $e_4$, bands to undergo impact ionization scattering by
creating electron hole pairs in the first and the second bands
respectively: $e^4_3=e^2_2+e^1_1+h^1_1$ and
$e^5_4=e^1_1+e^2_2+h^2_2$, where the upper subscript $L_i$ is the
angular momentum in units of $2/3d$. In addition, angular momentum
determines the band edge energy $\Delta_i$. In the large $d$ limit,
$\Delta_i=\hbar v_F L_i$, where $v_F$ is the Fermi velocity of
graphene. Therefore, in this limit, the energy and momentum
conservations are simultaneously satisfied for the electrons at the
bottom of the third and the fourth bands. However, for finite
diameter tubes: $\Delta_3<\Delta_2+2\Delta_1$ in $mod(n-m,3)=1$
chirality and $\Delta_4<\Delta_1+2\Delta_2$ in $mod(n-m,3)=-1$
chirality. This pushes the lowest impact ionization threshold energy
$E_{th3}$ of a (25, 0) CNT in the third band at energies higher than
its bottom $\Delta_3$, i. e. $E_{th3}>\Delta_3$,  and a power law
dependence is obtained $\Gamma_{ii3}\propto (E-E_{th3})^{\alpha}$
with $\alpha\approx 2$. Whereas at the bottom of the fourth band
$\Delta_4$, energy and momentum conservation laws are simultaneously
satisfied such that the threshold energy $E_{th4}$ coincides with
the bottom of the band $\Delta_4$, i. e. $E_{th4}=\Delta_4$. The
energy dependence of the scattering rate at the onset in the fourth
band is similar to that of a step function $\Gamma_{ii4}\propto
\Theta(E-E_{th4})$. In a (26, 0) tube the impact ionization
threshold in the third band coincides with $\Delta_3$, i. e.
$E_{th3}=\Delta_3$, and it is higher than $\Delta_4$ in the fourth
band, i. e. $E_{th4}>\Delta_4$. For impact ionization to take place
in the first (second) band via the decay channel
$e^1_1=e^1_1+e^1_1+h^{-1}_1$ ($e^2_2=e^2_2+e^1_1+h^{-1}_1$), the
threshold energy must be at least $E_{th1}\ge3\Delta_1$
($E_{th2}\ge\Delta_2+2\Delta_1$) and momentum conservation along the
1D wavevector in non-parabolic bands brings the threshold for the
first (second) bands to energies beyond the scale of
Fig.~\ref{fig1}a \cite{footnote1}.

In impact excitation, the exciton state formed has an effective mass
enhanced by 30\% over the free particle value ($m_e+m_h$), due to
the Coulomb interaction \cite{Perebeinos2}. (We find that the
magnitude of the  mass enhancement depends on the strength of
Couloumb interaction as $1/\varepsilon$, independent of tube
chirality.) This mass enhancement lowers the threshold energy.
Indeed, in the limit of the infinite effective mass, the exciton
dispersion is flat, and therefore, energy and angular momentum
conservations alone would determine the impact excitation threshold.
Thus, the impact excitation thresholds of the first two conduction
bands (Fig.~\ref{fig1}b) are significantly reduced from the impact
ionization values \cite{footnote1}. At the onset of impact
excitation, the third and the fourth band electrons can decay as
$e^4_3=e^2_2+E_{11}^2$ or $e^4_3=e^1_1+E_{12}^3$ and
$e^5_4=e^2_2+E_{12}^3$ or $e^5_4=e^1_1+E_{22}^4$. The relative
efficiencies of the two decay channels depend on tube chirality and
electron energy.

\begin{figure}
\includegraphics[height=2.4in]{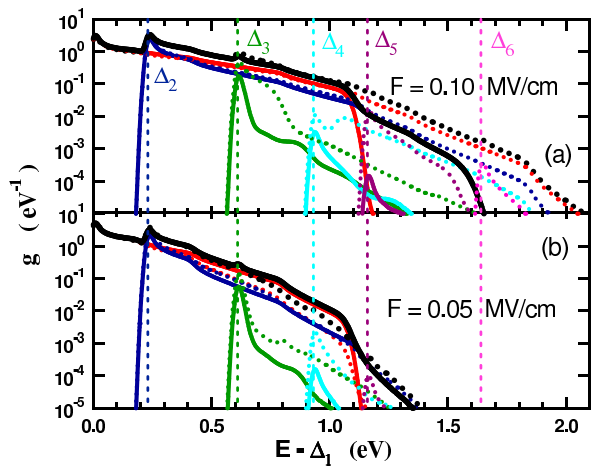}
\caption{\label{fig2}  Electron distribution function in a (25,0)
nanotube at applied electric field of (a) $F=0.1$ MV/cm and (b)
$F=0.05$ MV/cm as a function of energy in the first six conduction
bands shown in red, blue, green, cyan, magenta, and light red,
respectively. Black curves are the total distributions. In (a) solid
curves use impact excitation and dotted curves impact ionization
scattering rates with phonons at $T=300$ K. In (b) we use impact
excitation scattering, solid curves use phonon scattering at $T=300$
K and dotted curves use acoustic phonon scattering at $T_{ac}=300$ K
and optical phonon scattering at $T_{op}=1500$ K.}
\end{figure}

In electro-optical applications, we need to know the effect of
impact excitation scattering on the excited state production rate.
This depends on the hot carrier distribution function, which we
calculate by solving the steady-state multi-band Boltzmann equation
in the presence of an electric field, electron-phonon scattering
modeled as in \cite{Perebeinos3}, and either impact ionization
Eq.~(\ref{eq1}), or impact excitation scattering Eq.~(\ref{eq4}).
The results are shown in Fig.~\ref{fig2}. At energies above the
optical phonon energy of about $180$ meV, the carrier distribution
is determined by the field $F$ and the optical phonon mean free path
$\lambda_{op}$. The probability that an electron is accelerated to
an energy $E=eF\lambda$ over a length $\lambda$ without being
scattered by an optical phonon is
$\exp{\left(-\lambda/\lambda_{op}\right)}$. Therefore, the carrier
distribution $g$, for energies above the optical phonon and below
the impact excitation (ionization) threshold, is expected to follow
$g(E)\propto\exp{\left(-E/eF\lambda_{op}\right)}$, as seen in
Fig.~\ref{fig2}. Above the impact excitation (ionization) threshold,
the scattering is an order of magnitude larger and so is the slope
of $ln\left(g(E)\right)$ versus $E$, as in Fig.~\ref{fig2}. The
higher threshold for impact ionization results in a ``hotter"
carrier distribution than in the case of impact excitation
(Fig.~\ref{fig2}a).

The non-equilibrium phonon effect on the carrier distribution is
shown in Fig.~\ref{fig2}b. Although at optical phonon temperature
$T_{op}=1500$ K the phonon scattering rate is a factor 1.7 larger
than at room temperature, the slope of $ln\left(g(E)\right)$
versus $E$, for energies below the impact excitation threshold, is
similar to the $T_{op}=300$ K case. On the other hand, the hot
carrier tail above the impact excitation threshold significantly
increases with $T_{op}$, which in turn increases the exciton
production rate.

\begin{figure}
\includegraphics[height=2.4in]{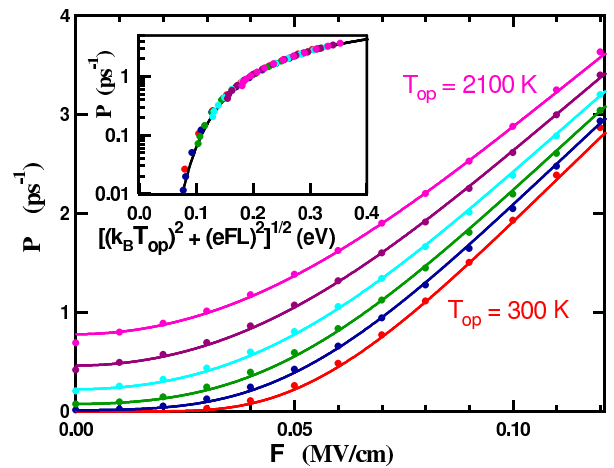}
\caption{\label{fig3}  Exciton production rate $P$ per unit carrier
in a (25,0) tube as function of the electric field for acoustic
phonon scattering at $T_{ac}=300$ K and different optical phonon
temperatures (from bottom to top) $T_{op}=$ 300 K (red), 900 K
(blue), 1200 K (green), 1500 K (cyan), 1800 K (magenta), 2100 K
(light red) along with the best fit to Eq.~(\protect{\ref{eq5}}) for
$\lambda_{op}=25$ nm, $E_t=0.57$ eV and $P_0=18$ ps$^{-1}$. The
inset shows scaling of all the calculations with the effective
temperature in accord with Eq.~(\protect{\ref{eq5}}).}
\end{figure}

The exciton production rate itself is given by the product of the
carrier distribution, as in Fig.~\ref{fig2}, and the impact
excitation scattering rate, as in Fig.~\ref{fig1}b, and it is shown
in Fig.~\ref{fig3}. We find that the rate $P$ can be well fitted by
the following equation:
\begin{eqnarray}
P=P_0\exp{\left(-\frac{E_t}{\sqrt{\left(k_BT_{op}\right)^2+\left(eF\lambda_{op}\right)^2}}\right)},
 \label{eq5}
\end{eqnarray}
where $E_t$ is the impact excitation threshold energy and $P_0$ is
a constant. The form of Eq.~(\ref{eq5}) resembles the Boltzmann
distribution with an effective temperature in the presence of the
field. The inset of Fig.~\ref{fig3} shows an excellent scaling of
the exciton production rate with the effective temperature. The
diameter dependence of the fit parameters in Eq.~(\ref{eq5}) for
$1$ nm $<d<2$ nm is  very simple: $E_{th}\approx 1.22 \ {(\rm eV \
nm)}/d$, $\lambda_{op}\approx 14 \times d$, while $P_0\approx 18$
ps$^{-1}$ is nearly diameter independent.

\begin{figure}
\includegraphics[height=3.4in]{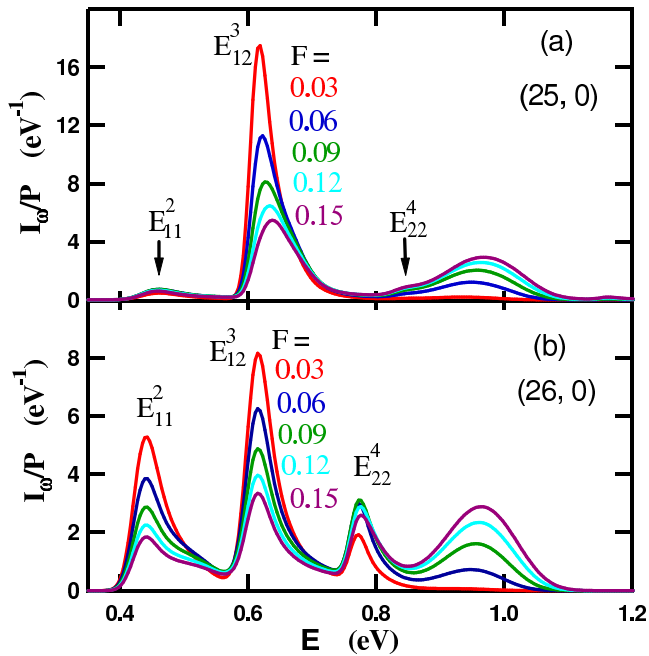}
\caption{\label{fig4}  Energy distribution $I_{\omega}$ of excitons
produced in the impact excitation process in (a) a (25,0) tube and
(b) a (26,0) tube with phonons at $T=300$ K and different applied
electric fields $F$ in MV/cm: 0.03 - red, 0.06 - blue, 0.09 - green,
0.12 - cyan, 0.15 - magenta, and normalized to the total production
rate $P=\int I_{\omega} d\omega$.}
\end{figure}

Excitons in nanotubes can be created either by optical pumping as
in photoluminescence experiments, or by electrical excitation as
in electroluminescence. In the photoluminescence case, the energy
of the created excitons follows the photon energy of the
excitation light source, and the angular momentum distribution is
determined by the dipole selection rules. Whereas in the impact
excitation process, the energy distribution of the generated
excitons depends on the applied bias and it is not subject to the
same selection rules. In Fig.~\ref{fig4} we plot the distribution
function of the initially created excitons at different fields.
The distribution has several distinct peaks, which we identify by
analyzing the contributions from excitons with different angular
momenta. At the onset of impact excitation, the third and the
fourth band electrons generate the low energy excitons with finite
angular momenta: $E^{2}_{11}$, $E^{3}_{12}$, $E^4_{22}$, whereas
at higher bias electrons in the first band produce unbound
excitons of zero angular momentum, which contribute significantly
to a broad continuum at energy of about 0.9 eV in Fig.~\ref{fig4}.

We conclude that CNTs and other 1D systems are characterized by
unusually high impact excitation rates and can be used as efficient
exciton sources to be employed in fundamental studies of interacting
boson systems at high densities and in opto-electronic device
applications.


\begin{thebibliography}{}
\bibitem{note1} Due to the mutual cancellation of the exciton
binding and the free particle $GW$ corrections \cite{Spataru}, we
use siccor approximation for the free particle spectra with a
siccor value equal to the binding energy of the first optically
active exciton.
\bibitem{footnote1} For two parabolic valence and conduction bands the
onset of the impact excitation has the usual power law dependence
\cite{Landsberg} $\Gamma\propto (E-E_{th})^{\alpha}$ with
$\alpha\approx 2$ and $E_{th}\approx 1.5 E_g$, where $E_g$ is the
bandgap. Due to the non-parabolicity of the CNT bands, in (25,0)
tube: $E_{th1}=\Delta_1+2.6$ eV, $E_{th2}=\Delta_1+4.6$ eV or
$E_{th}\approx 6.5 E_g$.
\end{thebibliography}
\end{document}